\documentstyle[epsf,12pt]{article}

\begin{document}
\renewcommand{\baselinestretch}{1.5}

\newcommand\beq{\begin{equation}}
\newcommand\eeq{\end{equation}}
\newcommand\bea{\begin{eqnarray}}
\newcommand\eea{\end{eqnarray}}

\newcommand\rRo{\rho_{R}^{A0}}
\newcommand\rLo{\rho_{L}^{A0}}
\newcommand\rR{\rho_{R}^A}
\newcommand\rL{\rho_{L}^A}

\newcommand\al{\alpha}
\newcommand {\dlt}{ \frac{\delta}{\pi}}
\newcommand {\dlts}{\frac{\delta^2}{\pi^2}}

\newcommand\pisig{\Pi_{\sigma i }}
\newcommand\pisigm{\Pi_{I\sigma i }}
\newcommand\sumsig{\sum_{\sigma i }}
\newcommand\psig{p_{\sigma i }}
\newcommand\xsig{x_{\sigma i }}
\newcommand\xmsig{x_{-\sigma j }}
\newcommand\sumI{\sum_{I}^{N}}

\newcommand\sumi{\sum_i}
\newcommand\sumj{\sum_j}
\newcommand\sumJ{\sum_J}

\newcommand\expo{e^{iS(\{x_{+i}\}, \{x_{-i}\})}}
\newcommand\expon{e^{-iS(\{x_{+i}\}, \{x_{-i}\})}}

\newcommand\trho{\tilde\rho}
\newcommand\tp{\tilde\phi}
\newcommand\dpi{\delta/\pi}
\newcommand\dtpi{\delta/2\pi}
\newcommand\ddpi{\delta^2/\pi^2}
\newcommand\px{\partial_x}
\newcommand\pt{\partial_t}
\newcommand\prl{Phys. Rev. Lett.}
\newcommand\prb{Phys. Rev. {\bf B}}
\hfill MRI-PHY/P981168
%\hfill cond-mat
%\vskip 0.5cm

\centerline{\bf Transport in Double-Crossed Luttinger Liquids}
\vskip 1 true cm

\centerline{P. Durganandini\footnote{{\it e-mail address}:
pdn@physics.unipune.ernet.in}}
\centerline{\it Department  of Physics, Pune University, }
\centerline{\it Pune 411 007,  India.}
\vskip .5 true cm

\centerline{Sumathi Rao \footnote{{\it e-mail
address}: srao@thwgs.cern.ch, sumathi@mri.ernet.in}}  
\centerline{\it Mehta Research Institute, Chhatnag Road, Jhunsi,}
\centerline{\it Allahabad 211 019, India.}

\vskip 2 true cm
\noindent {\bf Abstract}
\vskip 1 true cm
We study transport through two Luttinger liquids ( one-dimensional
electrons interacting through a Coulomb repulsion in a metal)
coupled together at {\it two}   points. External voltage biases are 
incorporated through boundary conditions. We include density-density
couplings  as well as single-particle hops at
the contacts. For weak repulsive interactions, transport through the
wires remains undisturbed by the inter-wire couplings, which renormalise
to zero. For strong repulsive interactions, the inter-wire couplings
become strong. For symmetric barriers  and no external voltage bias, 
a single gate voltage is
sufficient to tune for  resonance transmission in both wires. However, for
asymmetric couplings or for finite external biases, the system is
insulating.

\vskip 1 true cm

\noindent PACS numbers: 71.10.Pm, 72.10.-d, 73.40.Gk 

\newpage

Interest in the study of one-dimensional systems and Luttinger
liquid behaviour\cite{LUTT} has remained high ever since the discovery of high
$T_c$ superconductors and Anderson's suggestion \cite{ANDERSON}
that they could
be explained by two-dimensional Luttinger liquids. But 
the study of one-dimensional systems and Luttinger liquids 
has received a boost in recent years, since it was found that it
was actually possible to fabricate one-dimensional quantum wires
operating in the single channel limit\cite{1DWIRE}. 
Other experimentally realisable Luttinger liquid systems were the
edge states in the fractional quantum Hall bar\cite{EDGE}, 
carbon nanotubes\cite{CTUBES} and 
long chain molecules\cite{LCMOLS}.

Although still beset with contact problems, it has now become clear that 
transport measurements in such systems could directly probe Luttinger
liquid behaviour in the near future. A detailed paper by Kane and
Fisher\cite{KF} a few years ago  addressed the question of transport 
through wires with one or two weak barriers or, 
in the opposite limit, through one or two weak links.
This has led to a lot of further work in this area, with inclusion of 
external voltage biases\cite{EGGRAB,EG}, 
exact solutions at a particular value of the
Luttinger liquid coupling parameter\cite{GHALF} and 
finally the exact solution at
arbitrary coupling parameter using the thermodynamic 
Bethe ansatz\cite{LUDWIG}.

In a recent paper, Komnik and Egger\cite{KE} 
studied crossed Luttinger liquids - 
$i.e.$, two wires (modelled by spinless Luttinger liquids) coupled at 
a point. External potential biases were introduced through boundary
conditions\cite{EGGRAB} on the densities of the left and 
right-movers and inter-wire
density-density coupling and single particle hoppings were considered.
Interestingly, they found that the current through each wire was sensitive
to the cross-voltage ( the voltage drop across the other wire). This is
contrary to what is expected in an uncorrelated system and was hence
a sensitive test of Luttinger liquid behaviour.

In this paper, we extend this work and study doubly crossed Luttinger
liquids, or two wires coupled at {\it two} points. We also incorporate 
external biases through boundary conditions and consider density-density
and single particle hoppings. We find that for weak repulsive Coulomb
interactions, transport through both the wires remain undisturbed by
the inter-wire couplings. For strong repulsions, the inter-wire couplings
grow and become opaque to transmission. This fixes the charge on the
island between the two coupling points. Interestingly however, for
symmetric couplings between the wires, and no external voltage bias, a
single gate voltage is sufficient to tune for resonant 
transmission in both wires. For asymmetric couplings or with external
voltage biases, the system is insulating, except when the value of the
Luttinger liquid coupling is $g=1/2$. At this particular value of $g$,
even asymmetric couplings or finite external biases, lead to resonant
transmission, albeit less than perfect.

We start with two quantum wires with one-dimensional left and
right moving fermions. At the ends of the wire, we have external 
ideal reservoirs (a la Landauer), 
held at potentials $U_i^A$, $i=1,2$ at the left and right
edges and $A=1,2$ for the two wires.  ($U_1^A-U_2^A$ is the applied two 
terminal voltage in each wire.)
as shown in Fig.1. 
%%------------------CHANGE-BEGIN----------------------
\epsfxsize=3.8 in
\epsfysize=2.6 in
\begin{center}
\epsffile{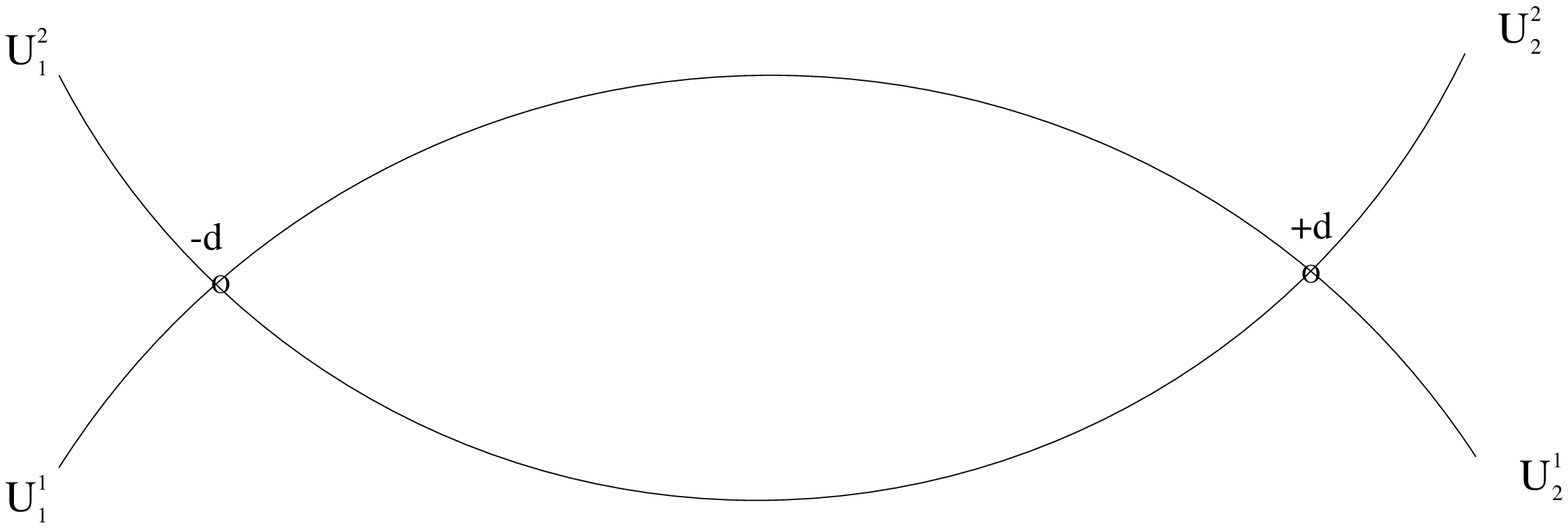}
\end{center}
\begin{itemize}
\item{ \bf Fig 1. } 
Two Luttinger liquids coupled together at two points ($x=-d$ and
$x=+d$) and connected to external reservoirs held at constant voltages
$U_1^1, U_1^2$ on the left and $U_2^1, U_2^2$ on the right.
\end{itemize}
%%------------------CHANGE-----END------------------

The two wires  are coupled together at two points located
at $-d$ and $d$. 
We introduce screening backgates around the quantum 
wire at a distance $R_s$, to screen the Coulomb interaction.
$R_s<<L$, where $L$ is the length of the wire, but is much larger
than the diameter of the quantum wire, which we take to be vanishingly
small.
  
We specialise to the case of spinless fermions and 
use the standard methods of abelian 
bosonisation (for a detailed comparison of different bosonisation
methods, see Ref.\cite{DS}) to write the Hamitonian
for each of the uncoupled wires in terms of a scalar field as
\beq
H = {\hbar v_F\over 2g} \sum_{A=1}^2
\int dx [g(\px \phi^A)^2 +g^{-1}(\px\theta^A)^2]
\eeq
where the fields $\phi^A$ and $\theta^A$  are 
related to the right and left-moving fermion fields 
as 
\beq
\psi_{p}^A(x) = {1\over (2\pi a)^{1/2}} {\rm exp}[-ipk_F x-i\sqrt{\pi}
\phi^A(x)
-ip\sqrt{\pi}\theta^A(x)]
\label{fermion}
\eeq
where $p=+/-$ stands for $R/L$ movers 
and $g$ is the Luttinger liquid parameter.
For future use, we note that the boson representation for the 
electron density is given by 
\beq
\rho^A (x) = {1\over \sqrt{\pi}} \px \theta^A (x) + {1\over \pi a}
\cos (2k_F x + 2\sqrt{\pi} \theta^A(x))
\label{density}
\eeq    
where we have dropped the equilibrium 
density $\rho=k_F/\pi$, (which is expected to be neutralised by positive
background charges.) The $2k_F$ oscillatory part of the density 
comes from the interference between the left- and right-movers. 
For a homogeneous wire, the interaction between the 'slow $\px
\theta$ mode' and the 'fast $2k_F$ mode' averages out and we only
need to take the slow modes into account. However, the oscillatory 
part can play a role in charging impurities.
 
To incorporate 
voltage biases at the ends of the wires, we need to put appropriate
boundary conditions on the fermion densities\cite{EG}. The left reservoirs
at $x=-L/2$, at  voltages $U_1^A$ inject right-moving charges 
\beq
\rho_R^{A0} (-L/2) = eU_1^A/2\pi \hbar v_F
\eeq
where $1/2\pi \hbar v_F$ is the density of states for right-moving particles,
and similarly the right reservoirs introduce left-moving charges, 
\beq
\rho_L^{A0} (L/2) = eU_2^A/2\pi \hbar v_F.
\eeq
(There is some subtlety in bosonisation for finite lengths, but we ignore
it under the assumption that $L$ is much larger than any other length
scale in the problem and can be effectively taken to be infinite.)
We assume that the outgoing particles enter the reservoirs with 
no reflection. In the presence of intra-wire interactions and inter-wire
couplings, the densities of  the right and left-moving 
particles in the wire are not fixed by the external biases, but rather
have to be  dynamically and self-consistently determined.

In the absence of any impurities and when there is no inter-wire
coupling,  the problem reduces to that of a single quantum wire for 
each of the wires. Hence, let us briefly review the properties of a 
single wire to set the notation. For a single wire with external
potentials $U_1$ and $U_2$, we can solve for the densities 
as follows\cite{EG}.
(All densities are measured with respect to the 
zero external voltage, non-interacting equilibrium 
density $\rho=k_F/\pi$.) 
Since the electrons are charged, and have Coulomb
repulsions, to increase the density of the electrons in the wire
through external potentials costs (Coulomb) energy.
This potential energy is neutralised beyond the screening length
by a screening backgate. (This can be a metal cylinder surrounding the
wire at a distance $R_s$.) But between the wire and the gate,
a capacitor is formed, whose charging energy is given by
\beq
{Q^2\over 2C} = {e^2 \over 2C}\int dx \rho^2(x)
\label{cap}
\eeq
This energy is provided by the interaction energy within the wire
given by
\beq
H_I = {1\over 2}\int dx dy\rho(x) U(x-y) \rho(y)
\eeq
where $U(x-y)$ is the Coulomb interaction, which in turn is 
related to the Luttinger liquid parameter $g$. As a concrete
example, let  us take the screened 
$U(x-y)$ to be of the form\cite{CHARGING}
\bea
U(x-y) &=& {U_0\over 2R_s} {\rm exp}(-|x-y|/R_s) \nonumber \\
\Longrightarrow  U_k &=& {U_0\over 1+R_sk^2}.
\eea 
The Luttinger liquid parameter $g$ is related to $U_0$ as
\beq
g=(1+U_0/\pi \hbar v_F)^{-1/2}.
\label{geq}
\eeq
For length scales larger than the screening length $R_s$,
the Coulomb interaction is just a $\delta$-function, so that
the interaction Hamiltonian becomes
\beq
H_I = {U_0\over 2} \int dx \rho^2(x)
\label{int}
\eeq
Comparing Eqs.(\ref{cap}) and (\ref{int}), we see that the 
capacitance per unit length is given by $C=e^2/U_0$.
The electrostatic potential drop $\phi(x)$ between the wire
and the screening gate is given by rewriting the interaction
energy as
\beq
H_I = {e\over 2} \int dx \rho(x) \phi(x)
\eeq
which gives $e\phi(x) = U_0 \rho(x)$.
For free fermions $U_0 = 0$ and there is no potential drop and
no need of screening.  

When we introduce a single impurity into the system, several effects
have to be taken into account. The impurity can act as a resistance 
impeding the flow of current. However, for an interacting system in
one dimension, the voltage drop across the impurity need not be 
the same as the voltage drop across the two ends of the wire. The
external voltage drop ( or the two-terminal voltage) is always given
by $U_1-U_2$. But the voltage drop across the impurity can range
from 0, for impurity  strength $\rightarrow 0$, to $(1-g^2) (U_1-U_2)$
for impurity strength $\rightarrow \infty$. Note that only for
$g\rightarrow 0$, the voltage drop across the impurity can be the
full two-terminal voltage. This is because for $g\rightarrow 0$, the
Coulomb interaction is so strong that there can be no excess charge on
the wire.

For our problem, let us first start with uncoupled wires.
If  $\rho_L^{A0}$ and $\rho_R^{A0}$ are introduced into the wire by the 
reservoirs, 
inclusion of the charging energy 
reduces the density of the particles in the quantum wire
by $e\phi \times$ density of states $1/\pi \hbar v_F$. (Remember that
the electrons are free to flow out through the outgoing reservoirs.)
So the
actual densities of the left and right movers in the quantum wires
satisfy the relation
\bea
\rho_R^A + \rho_L^A &=&\rho_R^{A0} + \rho_L^{A0} - e\phi/\pi \hbar
v_F\nonumber \\ &=& g^2(\rho_R^{A0} + \rho_L^{A0})
\eea
using Eq.(\ref{geq}). Since the difference of the densities of the 
right and left movers stays constant, ($\rho_R^A - \rho_L^A =
\rho_R^{A0} - \rho_L^{A0}$), we can solve 
for $\rRo$ and $\rLo$ in terms of $\rR$ and $\rL$ as
\bea
\rRo &=& {(1+g^2)\over 2g^2}\rR + {(1-g^2)\over 2g^2}\rL\nonumber \\
\rLo &=& {(1-g^2)\over 2g^2}\rR + {(1+g^2)\over 2g^2}\rL.
\eea
Hence,  we now 
see that the external voltages
can be written as boundary conditions for the $\theta^A$
fields\cite{EG},
\bea
({1\over g^2}\px + {1\over v_F}\pt)\theta^A(x=-L/2) &=& {eU_1^A\over \sqrt{\pi}
\hbar v_F}  \nonumber \\
({1\over g^2}\px - {1\over v_F}\pt)\theta^A(x=L/2) &=& {eU_2^A\over \sqrt{\pi}
\hbar v_F}.
\eea

Let us now consider
couplings between the two wires. Provided $d$ is large enough, the 
couplings at $-d$ and $d$ do not interfere. We can have density-
density couplings and single particle tunnelings
at the two contacts. For the case of a single contact, the
density-density and single-particle tunnelings have already been
studied for a single crossing in Ref.\cite{KE}. They
explicitly showed that for $g<1$, the effects of single-particle 
tunneling can be captured by renormalisation of the electro-static
density-density couplings. Hence, it is only necessary to keep the
density-density couplings. 
For the two crossing case, we have   
density-density electrostatic interactions at two points given by 
\beq
V_{\rm den} = \lambda_1 \rho^1(-d) \rho^2 (-d) +
             \lambda_2 \rho^1(d) \rho^2(d)
\eeq
Rewriting this interaction in terms of the phase fields using Eq.
(\ref{density}), (only the 'fast  modes') 
and keeping only terms with scaling dimension 
$\leq 1$,  we obtain
\beq
V_{\rm den}= \lambda_1 \cos (\sqrt{4\pi} \theta_1(-d)) \cos
(\sqrt{4\pi} \theta_2(-d)) + \lambda_2
\cos (\sqrt{4\pi} \theta_1(d)) \cos
(\sqrt{4\pi} \theta_2(d)) 
\eeq
with scaling dimension $\eta=2g$. ( $\eta=2g$ because rewriting the
Hamiltonian in the standard form with no interactions will require
a rescaling of the $\theta$ field by $\sqrt{g}$. )
All other terms are irrelevant and will renormalise to zero in
the low energy limit. Obviously, for $g<1/2$, this coupling is 
relevant and will grow at low energies. But unlike for scatterers
in a single wire, for $1/2<g<1$, this coupling is irrelevant for 
coupled wires.
Thus, for $g=1$, which is the appropriate limit for uncorrelated electrons, 
we obtain the results from the usual Landauer-Buttiker formalism 
in the geometry of Fig.(1).

For $1/2<g<1$  or weak repulsive interactions in the two wires, 
$\lambda_i \rightarrow 0$ explicitly, 
and there is perfect transmission independently in each wire 
at the fixed point. 
At zero temperature, the I-V characteristics are governed by the 
exact results given by 
\beq
I^{A} = {e^2\over h} U^{A}.
\eeq

What happens in the regime where the Luttinger interactions are strong?
Surprisingly,
for $0<g<1/2$ (for strong repulsive interactions),  the Hamiltonian
$H = H_0 +V_{\rm den}$ decouples in terms of the linear combinations
of the phase fields given by
\bea
\theta^{\pm} &=& (\theta^1\pm\theta^2)/\sqrt{2}\nonumber \\
{\rm and} \quad  \phi^{\pm} &=& (\phi^1\pm\phi^2)/\sqrt{2}
\eea
into a sum $H_{+} + H_{-}$ with
\beq
H_{\pm}= {\hbar v_F\over 2g} \int dx \{g(\px \phi_{\pm})^2 + g^{-1}(\px
\theta_{\pm})^2\}
\pm {\lambda_1\over 2} \cos[\sqrt{8\pi}\theta_{\pm}(-d)]
\pm {\lambda_2\over 2} \cos[\sqrt{8\pi}\theta_{\pm}(+d)]
\label{hami}
\eeq
and the boundary conditions on these fields are given by
\bea
({1\over g^2}\px + {1\over v_F}\pt)\theta^{\pm}(x=-L/2) &=& 
{(eU_1^1\pm eU_1^2)\over \sqrt{\pi} \hbar
v_F}  \nonumber \\
({1\over g^2}\px - {1\over v_F}\pt)\theta^{\pm}(x=L/2) &=& 
{(eU_2^1\pm eU_2^2)\over \sqrt{\pi} \hbar
v_F}
\label{bc}
\eea

Hence, we have two decoupled Hamiltonions, each with Luttinger
interactions $\tilde g =2g$ (because rescaling the kinetic term gives
the dimension of the barrier terms as $\eta=2g=\tilde g$ and we do not
rescale the sound velocity $v=v_F/g$) and each with two
scatterers at the positions $\pm d$. So just like the single 
crossing case considered by Komnik and Egger\cite{KE}, the
doubly crossed Luttinger liquid also gets mapped into two single wire
problems (albeit with twice the interaction strength) and 
with two potential scatterers. 

The case for $g=1/2$, ($\tilde g=1$) can now be trivially
addressed. It corresponds to a system of totally uncorrelated fermions
with two barriers at $\pm d$. For a single barrier, it is easy to see
by refermionising that the transmission coefficient is given by
\beq
T = {1\over 1+(\lambda/2)^2}.
\eeq
This agrees with the result in Ref.\cite{KE} that for a single
barrier, the $I-V$ characteristic for the two wires is given by 
\beq
I^{A} = {e^2\over h} T U^A = {e^2\over h(1+(\lambda/2)^2)} U^{A}
\eeq
For two barriers, the transmission coefficient depends on whether the
scatterers are combined coherently or incoherently\cite{DATTA}. For
incoherent scattering, 
\beq
T(E)= {T_1 T_2\over 1-2\sqrt{R_1 R_2} \cos\theta (E) +R_1 R_2}
\eeq
where $E$ is the incident energy, $\theta(E)$ is the phase shift
acquired in a round trip between the barriers , $T_i ={1\over 
(1+(\lambda_i/2)^2)}$ are the two transmission coefficients through the two
barriers and $R_i=1-T_i$ are the reflection coefficients.
Clearly, depending on the incident energies, the transmission varies
and resonant transmission is possible, whenever the denominator
becomes small. For incoherent scattering, 
\beq
T(E)= {T_1 T_2\over 1-R_1 R_2}
\eeq
which is independent of the phase $\theta$. Here, there exists no
possibility of resonances.

An interesting point to note in the above analysis is that resonant
transmission is possible, both for symmetric barriers ( in which case
$T_1 = T_2$ and the transmission is perfect, T($E_{\rm res}$) = 1),
and for asymmetric barriers, where the transmission is less than perfect.

For the correlated system, ($g\ne 1/2$), this problem 
(without the external potentials) was originally addressed
in the pioneering Kane-Fisher\cite{KF} paper, where it was shown that 
although in the strong interaction regime, 
the scattering potentials grow at low energies,
and prevent transport, for two scatterers, it was possible to
have resonant scattering by tuning a single parameter, either 
the incident energy or  a backgate voltage. 
In the large barrier limit, the charge on the island between the
two barriers is fixed and there will be an energy barrier to add
another electron. However, by tuning a gate voltage, we can make
the energy cost to add another electron vanish. Hence, as a function 
of the gate voltage, one can get resonances.
The resonance condition is fixed by having $E(n) = E(n+1)$, where
$E(n)$ is the energy of the system with charge $ne$ on the island
and is given by
\beq
4eg^2\Delta \phi_G^{\pm} = e{\tilde g}^2 \Delta \phi_g^{\pm} = \pi
\hbar v_F/2d
\eeq
where $\Delta \phi_G^{\pm}$ is the change in the gate voltage required to
move from one resonance to the other for the $\pm$ wires.
However, in this picture, the charging of the impurity was ignored.
Later work\cite{CHARGING} included the effects of charging the impurity.
For a single impurity, this effect was not observable, because
of the large capacitance between the wire and the screening
backgate. However, for two impurities, inclusion of these effects
changes the resonance condition, which becomes\cite{CHARGING}
\beq
4eg^2\Delta \phi_G^{\pm} = (2d/\pi \hbar v_F+ 2C^{\pm}(2g)/e^2)^{-1}
\eeq
where $C^{\pm}(2g)=C_{fr}^{\pm}(2g)+C_{2k_F}^{\pm}(2g)$. 
$C_{fr}^{\pm} = Q/U$ is the finite range capacitance due to
the total antisymmetric charge density on the wire in the presence of the
impurity. (The symmetric part of the charge density exists even in the absence
of the impurity and as explained earlier is due to imperfect
screening.) $C_{2k_F}^{\pm}$, on the other hand, comes from the $2k_F$
component of $\rho$ in Eq.(\ref{density}) 
and can be finite even for a zero range
($R_s = 0$) Luttinger liquid. The total capacitance is the sum of the
two contributions. We see that inclusion of the contribution of
the capacitance decreases the 
spacing of the gate voltage at which one gets resonances.
Note that the capacitances depend on $2g$ now
instead of $g$ for a single wire. Hence, the resonance spacing changes
from that of the single wire. However, since there is no dependence on
the external potentials, $C^{+}=C^{-}$. 
Hence, $\Delta\phi_G^+ = \Delta\phi_G^- = \Delta\phi_G$, 
which leads to the following very interesting consequence, 
for coupled wires with symmetric inter-wire couplings and 
zero external bias,  -
\bea
\Delta\phi_G^1 &=& \Delta\phi_G^+ + \Delta\phi_G^- = 2\Delta \phi_G \\
\Delta\phi_G^2 &=& \Delta\phi_G^+ - \Delta\phi_G^- = 0
\label{result}
\eea
-, $i.e.$, a single gate voltage is sufficient to have resonant
transmission in both wires.

What happens when we include 
finite external
potential differences between the two ends of the quantum wire?
The correct way\cite{DRAO} to address the problem is to 
solve the Hamiltonian with
two scatterers given in Eq.(\ref{hami}) with the boundary conditions in
Eq.(\ref{bc}). However, we may use the argument that
the effect of the external
bias serves to tilt the potential between the barriers. Hence, even if
we originally start with symmetric barriers, the external potential bias makes
them asymmetric. Thus, the problem of a quantum wire with external
potential bias and two scatterers, is analagous to the problem without
external bias, but with asymmetric scatterers. In this case, it is not
possible to get a zero-energy bound state, which was what was needed to
tune the resonance for symmetric barriers. Thus,
using the argument in Ref.\cite{KF} we see that, at least for zero
temperature, for coupled wires with an external bias, 
there is no resonance, and the system is
insulating. 

For weak barriers, even the capacitances do have dependences on the
external potentials. Moreover, the strong barrier assumption that the
potential drop across each barrier is equal to the two-terminal voltage
across the barrier, that was used to derive the formulae for $C_{fr}$
and $C_{2k_F}$ in Refs.\cite{CHARGING} and \cite{EGGRAB} is no longer valid. 
The explicit four terminal voltage drops across the barriers $V_1$ and
$V_2$ have to be computed\cite{EG} and used in calculating the
capacitances, but that is beyond the scope of this paper\cite{DRAO}.

We can generalise this result
to the case of $N$ crossings of two
wires. It is easy to see that the same definition of the $+$ and $-$
channels as the sum and difference of the fields in the two wires,
maps the model to two decoupled wires with coupling strength $2g$ and
$N$ potential scatterers. Naively, one may expect that it would be
hard to get resonances, since the scattering at each potential would
be incoherent. Hence, the system should be opaque to
transmission. However, if the crossings are perfectly symmetric and 
periodic, we expect the
equivalent of Bloch states and complete conductivity for specific
incident energies or equivalently, specific
tuned values of a single gate voltage\cite{DRAO}.

The other generalisation that one can consider is what happens if
three or more wires meet at a point. Here, again, if we assume
density-density couplings at a point, a naive
generalisation of the interaction term shows that the interaction
potential becomes irrelevant except when $3g<1$, or more generally
when $m$ wires meet a point, except when $mg<1$. For this very strong
interaction region, a similar analysis can be attempted. However, the
simplification of mapping the model to two decoupled wires with
interactions does not go through. Moreover, this geometry is less
physically relevant, unless it is possible to fabricate several wires
to make contact at a point.

To conclude, we emphasize the new results in this paper. We
have shown that transport in two wires crossing at two points remains
unaffected by the crossings for $g>1/2$. This is an extension of the
earlier result for single crossing in Ref.\cite{KE}. 
For $g<1/2$, we map the
problem to two decoupled wires with two impurities, again following
Ref.\cite{KE}, where it was done for a single crossing. For the decoupled
wires, we show that the resonance condition changes, since the
coupling strength is now $2g$ instead of $g$. The $g=1/2$ case has
uncorrelated fermions in the two wires and the standard analysis for resonance
transmission through a double barrier applies\cite{DATTA}. 
For symmetrically coupled unbiased 
wires with correlated electrons, ($g < 1/2$),  
we show that a single gate voltage suffices to tune the resonance 
condition in both the wires and have maximal transmission.
However, for asymmetric couplings between the wires or for wires with
external voltages, there is no transmission, except when $g=1/2$, in
which case, there does exist resonant transmission, although not
perfect. Finally, we
argue that for symmetric crossings at periodic lengths, 
it is still possible to
have resonant tunneling tuned by a single gate voltage.

\section*{Acknowledgments}
PD thanks C.S.I.R. (India) for financial support. 
Both of us thank the Abdus Salam 
ICTP for hospitality during the course of this
work. We would also like to thank the workshop held at Trieste 
on 'Strongly Correlated
Electron Systems' for providing the inspiration.

\end{document}